\documentclass[aps,prl,twocolumn,groupedaddress,showpacs,floatfix,preprintnumbers]{revtex4-1}

\usepackage{amssymb}
\usepackage{color,graphicx}

\begin{document}

\title{Dynamics and Interaction of Vortex Lines in an Elongated Bose-Einstein Condensate}

\author{S.~Serafini$^1$}
\author{M.~Barbiero$^1$}
\author{M.~Debortoli$^1$}
\author{S.~Donadello$^{1,2}$}
\author{F.~Larcher$^1$}
\author{F.~Dalfovo$^{1,2}$}
\author{G.~Lamporesi$^{1,2}$}
\email[]{giacomo.lamporesi@ino.it}
\author{G.~Ferrari$^{1,2}$}

\affiliation{\textit{1 INO-CNR BEC Center and Dipartimento di Fisica, Universit\`a di Trento, 38123 Povo, Italy}\\
\textit{2 Trento Institute for Fundamental Physics and Applications, INFN, 38123 Povo, Italy}}
\date{\today}

\begin{abstract}
We study the real-time dynamics of vortices in a large elongated Bose-Einstein condensate (BEC) of sodium atoms using a stroboscopic technique. Vortices are produced via the Kibble-Zurek mechanism in a quench across the BEC transition and they slowly precess keeping their orientation perpendicular to the long axis of the trap as expected for solitonic vortices in a highly anisotropic condensate. Good agreement with theoretical predictions is found for the precession period as a function of the orbit amplitude and the number of condensed atoms. In configurations with two or more vortices, we see signatures of vortex-vortex interaction in the shape and visibility of the orbits. In addition, when more than two vortices are present, their decay is faster than the thermal decay observed for one or two vortices. The possible role of vortex reconnection processes is discussed.

\end{abstract}

\pacs{03.75.Lm, 67.85.De, 05.30.Jp}

\maketitle

Vortex dynamics is an essential feature of quantum fluids \cite{Feynman55} and plays a key role in superfluid helium \cite{Donnelly91}, superconductors \cite{Tinkham96}, neutron stars \cite{Baym75} and magnetohydrodynamics \cite{Lalescu15}. The interaction between vortices is crucial for understanding the formation of vortex lattices in rotating superfluids and is the basic mechanism leading to quantum turbulence {\it via} vortex reconnection
\cite{Paoletti11,Zuccher12}. Vortices have been extensively investigated in atomic gases \cite{Fetter09}, where a variety of techniques permits the observation of single ones up to a few hundreds, interacting in a clean environment and on a spatial scale ranging from the healing length (core size) $\xi$  to a few tens of $\xi$. The fact that atoms are confined by external fields of tunable geometry makes them suitable to explore the physics of reconnection and dissipation in inhomogeneous
systems and in the presence of boundaries. Seminal experiments were performed in rotating Bose-Einstein condensates (BECs), where the effect of rotation and long-range interaction favors vortex alignment and the formation of vortex lattices \cite{Madison00,AboShaeer01,Engels02,Engels03,Coddington03} and hence crossing and reconnection mechanisms are inhibited. Interacting vortices have been observed in nonrotating oblate BECs, where vortex lines are short and either parallel or antiparallel, thus behaving as pointlike
particles dominated by their long-range interaction in a quasi-2D background \cite{Weiler08,Neely10,Freilich10,Middelkamp11,Navarro13,Kwon14}. 

In our experiment we use a cigar-shaped BEC  which is particularly suitable for studying the dynamics of vortex lines in 3D. Because of the boundary conditions imposed by the tight radial confinement each vortex line lies in a plane perpendicular to the long axis $z$ of the trap, such to minimize its length and therefore its energy, as in the solitonic vortex configuration predicted in Refs. \cite{Brand02,Komineas03} and recently observed both in a BEC \cite{Donadello14,Tylutki15} and in a superfluid Fermi gas \cite{Ku14}. The line is randomly oriented in the plane and away from it, at distances of the order of the system transverse size, the superfluid flow quickly vanishes and the long-range part of the vortex-vortex interaction is suppressed. Hence, vortices can move almost independently along elliptic orbits except when they approach each other and may collide with a random relative angle. At the scale of the healing length, where reconnection can take place, the system is still equivalent to a uniform superfluid, like liquid He, but with the advantage that vortex filaments collide at measurable relative velocities. 

\begin{figure}[!t]       %--------------FIGURE 1--------------
\centering
\includegraphics[width=\columnwidth]{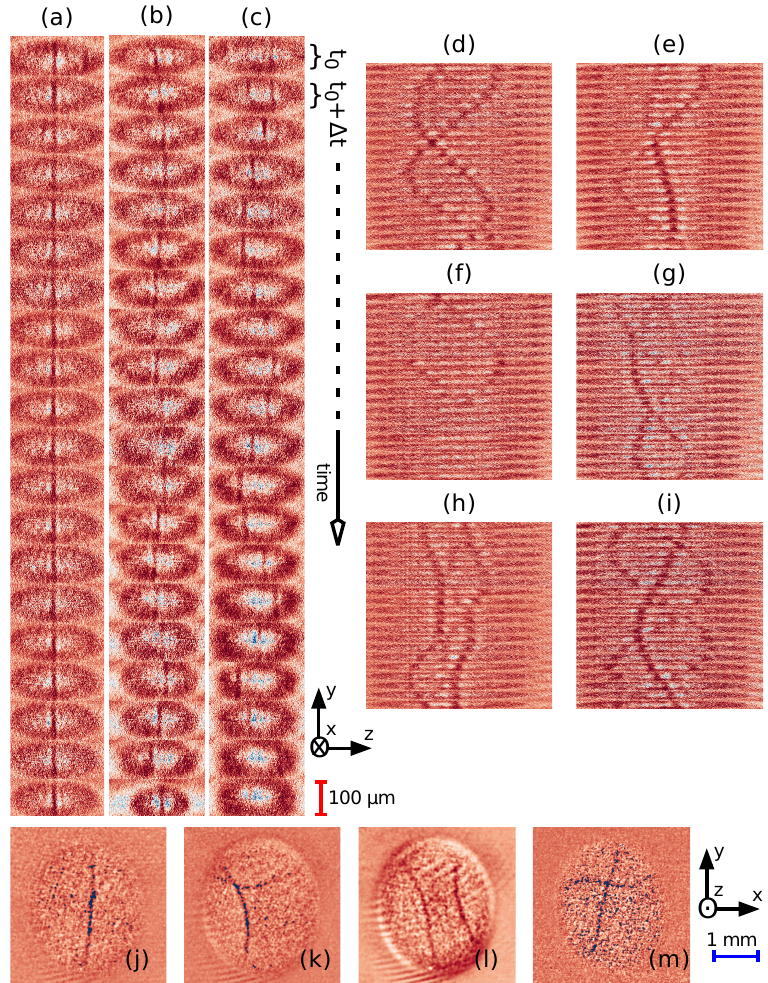}
\caption{(a)-(c) Sequences of $20$ images of the density distribution of the atoms extracted from three BECs; frames are taken every $\Delta t=84$~ms, each after a $13$ ms expansion. (a) Static vortex. (b)-(c) Vortices precessing with different amplitudes. Each vortex is randomly oriented in the $xy$ plane and, after expansion, it forms a planar density depletion \cite{Tylutki15} which is visible as a stripe. (d)-(i) Sequences with two and three vortices, with $\Delta t=28$~ms; here frames are not to scale and vertically squeezed to enhance visibility. (j)-(m) Destructive absorption images of the whole BEC taken along the axial direction $z$ after $120$~ms of expansion, showing (j) a single vortex filament crossing the condensate from side to side and (k)-(m) two vortices with different relative orientation and shape. All images show the residuals after subtracting the fitting TF profile.
}
\label{Figure1}
\end{figure}

The experimental apparatus is described in Ref. \cite{Lamporesi13}. We evaporate sodium atoms in a magnetic harmonic trap with frequencies $\{\omega_{x,y}=\omega_{\perp},\omega_z\}/2\pi=\{131,13\}$~Hz. Vortices with random position and velocity spontaneously originate {\it via} the Kibble-Zurek mechanism \cite{Kibble80,Zurek85,Weiler08,LamporesiKZM13} from phase defects in the condensate when crossing the BEC transition and their average number scales as a power law with the evaporation rate. At the end of the evaporation we have an almost pure prolate BEC with about $10^7$ atoms at $200$~nK in the state $|F,m_F\rangle=|1,-1\rangle$. In Refs. \cite{LamporesiKZM13,Donadello14} we counted and characterized defects using destructive absorption imaging. Here we apply a stroboscopic technique, similar to that in Refs. \cite{Freilich10,Ramanathan12}, which allows us to observe the real-time dynamics. Starting from an initial number of atoms $N_0$, we remove a small fraction $\Delta N/N_0\sim 4\%$ by outcoupling them to the antitrapped state $|2,-2\rangle$ {\it via} a microwave pulse, short enough to provide a resonance condition throughout the whole sample.
Outcoupled atoms are imaged along a radial direction after a $13$~ms expansion \cite{SM} without affecting the trapped ones. The extraction mechanism is repeated $20$ times with time steps $\Delta t$, keeping $\Delta N$ fixed. Raw images are fitted to a Thomas-Fermi (TF) profile \cite{Dalfovo99} and the residuals are calculated. Because of the peculiar structure of the superfluid flow of solitonic vortices \cite{Donadello14,Tylutki15}, after expansion the whole radial plane containing a vortex
exhibits a density depletion and vortices are seen as dark stripes independently of their in-plane orientation. During the extraction sequence the remaining condensate evolves in trap, only weakly affected by atom number change, provided $\Delta N/N(t)$ is sufficiently small. We can then identify the axial position of the vortex in each image of the outcoupled atoms and analyze its oscillation as a faithful representation of the in-trap dynamics. Typical examples are shown in Figs.~\ref{Figure1}(a)-\ref{Figure1}(i) . Alternatively we image the full BEC along the axial direction after a long expansion with a destructive technique as in \cite{Donadello14} and directly see the shape and orientation of the vortex lines as in Figs.~\ref{Figure1}(j)-\ref{Figure1}(m).

\begin{figure*}[tbp]       %--------------FIGURE 2--------------
\centering
\includegraphics[width=2\columnwidth]{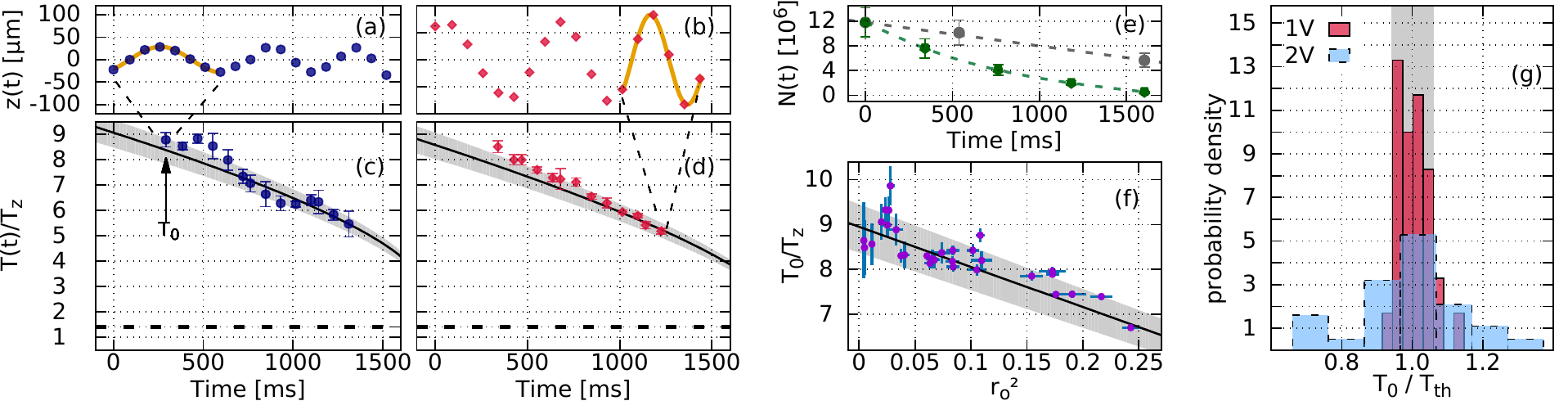}
\caption{(a),(b) Vortex axial position after expansion for the condensates in Figs.~\ref{Figure1}(b) and \ref{Figure1}(c). (c),(d) Instantaneous period normalized to the trapping period $T_z = 77$~ms (points) obtained by fitting the above oscillations; the solid line is the theoretical prediction (\ref{eqn:period}) for the measured atom number $N(t)$ and its $20\%$ uncertainty (grey region); the dashed line is the prediction for a dark or grey soliton. (e) BEC atom number, with (green) and without (grey) the extraction sequence. (f) Period $T$ extracted from the vortex position in the first frames in units of $T_z$ as a function of $r_{\rm o}^2$; the solid line represents the predicted $(1-r_{\rm o}^2)$ behavior, with no free parameters. (g) Probability density of the measured period $T_0$ {\it vs.} the theoretical one $T_{\rm th}$ in the same conditions. Red (blue) bars refer to $30$ ($27$) cases with a single  vortex (two vortices), all of them with the same $N$ within a $20\%$ uncertainty. 
}
\label{Figure2}
\end{figure*}

We first choose an evaporation rate of $525$~kHz/s, yielding one vortex in each BEC on average.  From the sequence of radial images we extract the axial position of each vortex $z(t)$. Frames are recorded every $\Delta t=84$~ms. Figures~\ref{Figure2}(a) and \ref{Figure2}(b) show two examples corresponding to the raw images of Figs.~\ref{Figure1}(b) and \ref{Figure1}(c), respectively. The observations are consistent with a vortex precession around the trap center, as the one observed in oblate BECs \cite{Anderson2000,Freilich10}. In a nonrotating elongated condensate, a straight vortex line, oriented in a radial plane, is expected to follow an elliptic orbit in a plane orthogonal to the vortex line, corresponding to a trajectory at constant density \cite{Sheehy04}. The observed motion of each dark stripe in Figs.~\ref{Figure1}(a)-\ref{Figure1}(c) is the axial projection of such a precession. Given $r_{\rm o}=z_{\rm max}/R_z=y_{\rm max}/R_\perp$ the in-trap amplitude of the orbit normalized to the TF radii $R_\perp= \sqrt{2\mu/(m\omega_\perp^2)}$ and $R_z= \sqrt{2\mu/(m\omega_z^2)}$ \cite{Dalfovo99}, the precession period is predicted to be   
\begin{equation}
T = \frac{4(1-r_{\rm o}^2) \mu}{3\hbar\omega_\perp \ln(R_\perp /\xi)} T_z\, , 
\label{eqn:period}
\end{equation}
where $T_z=2\pi/\omega_z$ is the axial trapping period and $\xi$ is related to the chemical potential $\mu$ by $\xi=\sqrt{\hbar^2/(2m\mu)}$. This result, which is valid to logarithmic accuracy, has been derived for a disk-shaped nonaxisymmetric condensate in Refs.~\cite{Svidzinsky00,Fetter01} within the Gross-Pitaevskii theory at  $T=0$ and in the TF approximation, corresponding to $R_\perp /\xi \gg 1$ (in our case, $R_\perp /\xi$ ranges from $60$ to $20$). It can also be obtained by means of
the superfluid hydrodynamic approach introduced in Ref. \cite{Pitaevskii13} to describe the motion of vortex rings in elongated condensates, appropriately generalized to the case of solitonic vortices as in Ref. \cite{Ku14}. The quantity $\mu(1-r_{\rm o}^2)$ is the local chemical potential along the vortex trajectory and we assume $r_{\rm o}$ to be constant during expansion, as distances are expected to scale in the same way in the slow axial expansion. 

In comparing the observed period with Eq.~(\ref{eqn:period}) we must consider that the number of atoms is decreasing from shot to shot.  Since extraction is spatially homogeneous, the gradients of the density, and hence the equipotential lines for the vortex precession and the orbit amplitude remain almost unchanged. 
However, $N(t)$ (hence $\mu \propto N^{2/5}$) decreases in time and so does the vortex orbital period $T$, as is clearly visible in Figs.~\ref{Figure2}(a) and \ref{Figure2}(b).
We define an instantaneous period at time $t$ as the period obtained from a sinusoidal fit to the measured position in a time interval centered at $t$ and containing about one oscillation. Such $T(t)$ is plotted in Fig.~\ref{Figure2}(c) and \ref{Figure2}(d) and compared to Eq.~(\ref{eqn:period}), where we include the effect of the observed $t$ dependence on $N$, shown in Fig.~\ref{Figure2}(e), both in $\mu$ and $\xi$. The agreement is good, the major limitation being the experimental uncertainty in $N$. We also show the period expected for the oscillation of a dark or grey soliton, which is $\sqrt{2}\ T_z$ independently of $N$ \cite{Busch00,Konotop04}. In Fig.~\ref{Figure2}(f) we plot the period of vortices orbiting with different amplitude $r_{\rm o}$. The agreement with theory is again good and can be further appreciated by considering the ratio between each value of $T$ measured at a given $r_{\rm o}$ and the theoretical value in Eq.~(\ref{eqn:period}) obtained for the same $r_{\rm o}$ and $N$. Figure~\ref{Figure2}(g) shows the histogram of all values obtained by extracting $T$ and $r_{\rm o}$ from a fit to the first oscillation, using $N=9\times 10^6$ in Eq.~(\ref{eqn:period}). The histogram gives $T_0/T_{th}=0.97\pm 0.04$. This remarkable agreement with theory is nontrivial since Eq.~(\ref{eqn:period}) assumes $r_{\rm o}\ll 1$ and a rigid straight vortex line, while off-centered vortices actually bend toward the curved BEC surface. For rotating condensates the bending mechanism has been discussed in Refs. \cite{Svidzinsky00b,Aftalion01,Garcia01a,Garcia01b,Modugno03} and observed in Ref. \cite{Rosenbusch02}. Examples of straight and bent vortices in our condensate are given in Figs.~\ref{Figure1}(j)-\ref{Figure1}(m). In our elongated BEC, with strong radial inhomogeneity, this bending mechanism is expected to be more effective than in oblate BECs. Our observations seem to indicate that its effect on the period is small, possibly of the same order of the logarithmic corrections to Eq.~(\ref{eqn:period}) predicted for a straight vortex in a 2D geometry \cite{Lundh00,Kim04}. This may be due to the fact that the difference in length between a bent and a straight vortex, at a comparable $r_{\rm o}$, is relatively small and the overall structure of the vortical flow is also quite similar, so that the key quantities entering the hydrodynamic description (i.e, the force acting on a unit of length of the vortex and the momentum of the vortex, in the language of Ref.~\cite{Pitaevskii13}) are almost the same in the two cases.

\begin{figure}[b!]       %--------------FIGURE 3--------------
\centering
\includegraphics[width=\columnwidth]{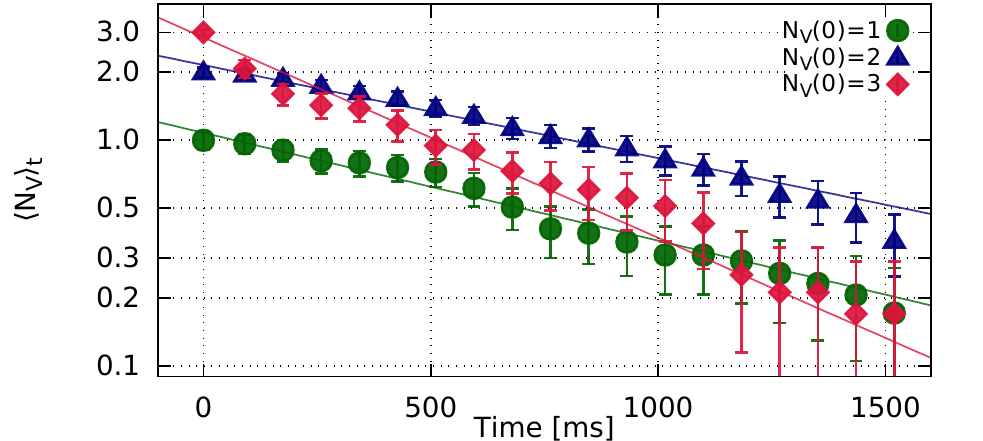}
\caption{Average vortex number, $\langle N_V\rangle$, remaining in a condensate at time $t$ starting from configurations with $N_V=1$ (circles), $2$ (triangles) and $3$ (diamonds) at $t=0$. Solid lines are exponential fits.}
\label{Figure3}
\end{figure} 

Vortex lifetime in nonrotating BECs is limited by scattering of thermal excitations, which causes the dissipation of the vortex energy into the thermal cloud. Since a vortex behaves as a particle of negative mass, dissipation causes an antidamping of the orbital motion and vortices decay at the edge of the condensate \cite{Fedichev99,Yefsah13}. We can measure the lifetime $\tau$ by counting the average number of vortices $\langle N_V\rangle_t$ remaining in the condensate at time $t$, starting
with $N_V(0)$. If $N_V(0)=1$ we find a clear exponential decay with $\tau_1=(910 \pm 100)$ ms (Fig.~\ref{Figure3}), close to that measured in Refs. \cite{LamporesiKZM13,Donadello14} and of the same order of the one observed in a fermionic superfluid \cite{Yefsah13,Ku14}.

\begin{figure}[tbp]       %--------------FIGURE 4--------------
\centering
\includegraphics[width=\columnwidth]{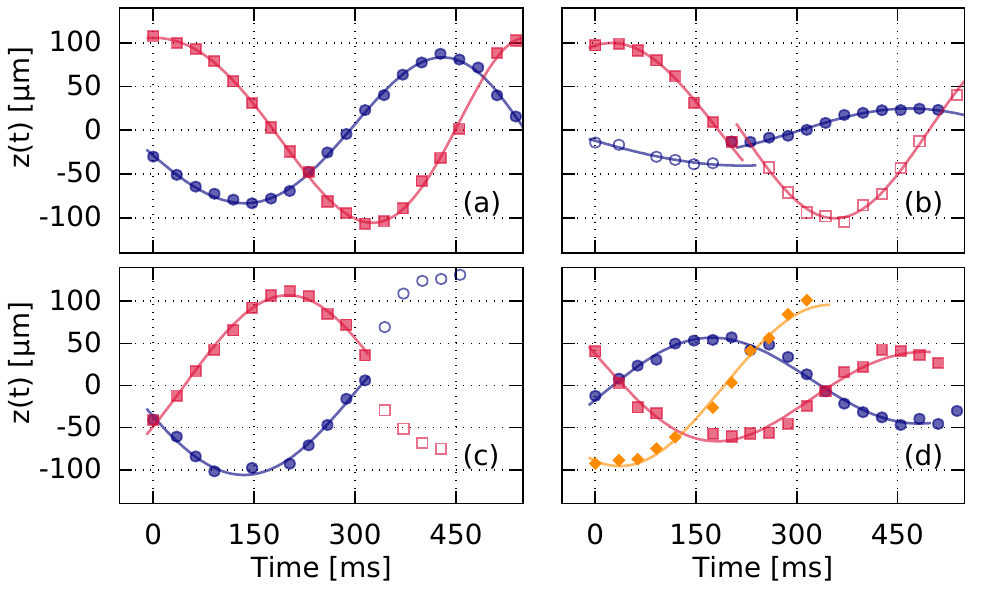}
\caption{Vortex axial position in BECs.  (a) Two vortices with no apparent interaction. (b) Two crossing vortices change their visibility and experience a phase shift in their trajectory. (c) Two vortices becoming hardly visible after crossing. (d) Two vortices oscillating with unperturbed trajectories while a third one disappears. (a)-(d) correspond to the data in Figs.~\ref{Figure1}(d)-\ref{Figure1}(g), respectively. Solid and empty symbols are used to distinguish high and low density contrast, respectively. }
\label{Figure4}
\end{figure}

Using a faster evaporation ramp ($700$ kHz/s), we produce more vortices and search for signatures of mutual interaction. Examples are shown in Fig.~\ref{Figure1}(d)-\ref{Figure1}(i) and typical trajectories are also reported in Fig.~\ref{Figure4}. In some cases, vortices perform unperturbed oscillations [Fig.~\ref{Figure4}(a)]; in others, we clearly see a shift in their trajectories at the crossing point [Fig.~\ref{Figure4}(b)]. The average relative velocity at the crossing in the latter case is systematically
smaller ($\sim0.5$ mm/s) than in the former ($\sim1.1$ mm/s) \cite{SM}. The shift has a consequence also in the determination of the orbital period as it causes a broadening of the probability distribution of the ratio $T_0/T_{th}$ which now gives $0.96\pm 0.14$, with a standard deviation three times larger than for the single vortex [Fig.~\ref{Figure2}(g)]. In addition, crossings are frequently associated with a sudden change of visibility of one or both vortices [Figs.~\ref{Figure1}(e)-\ref{Figure1}(h)). Finally, by analyzing the lifetime of vortices for the initial condition $N_V=2$ and $N_V=3$ we observe a lifetime $\tau_2=(1050 \pm 100)$~ms for the two-vortex configuration, consistent with the one-vortex configuration. The situation instead changes in the three-vortex configuration, where a faster decay is observed, $\tau_3=(490\pm100)$ ms (Fig.~\ref{Figure3}).

The frequent observation of unperturbed orbits for multiple vortices is intriguing. Two vortex lines moving back and forth in the condensate with random radial orientations should have large probability to cross each other at some point. If crossings occur, reconnections are expected to take place \cite{Zuccher12} with possible drastic (and almost temperature independent \cite{Paoletti10,Allen14}) effects on the vortical dynamics. The actual dynamics can strongly depend on the relative angle
$\alpha$ between vortex lines as well as the relative velocity $v_r$ between the planes where they lie. When $\alpha$ is close to $0$ ($\pi$), the vortex lines tend to align (antialign), thus reducing the chance of reconnection for vortices on different orbits. But when vortices approach with $\alpha\sim\pi/2$ reconnection can be hardly avoided. The fact that we observe the same vortex lifetime for $N_V(0)=1$ and $2$ implies that such reconnections are either suppressed or they
induce a negligible dissipation. A possible explanation is the occurrence of double reconnection processes \cite{Berry12}. Vortex reconnection corresponds to the switching of a pair of locally coplanar vortex lines, accompanied by a change of topology. In our geometry a finite $v_r$ implies that the newly formed filaments must stretch in the condensate while the two planes separate again after reconnection. The consequent energy cost is instead avoided if vortices perform a consecutive second
reconnection when they are still at close distance. This would preserve the vortex number, consistent with our observation of an equal vortex lifetime for $N_V(0)=1$ and $2$. It is worth mentioning that a similar scenario has also been recently suggested for the collision of cosmic strings \cite{Verbiest11}. The occurrence of a shift in the trajectories, that apparently depends on $v_r$, could be associated with the role of the collision time: faster vortices have less time to interact and their
trajectories are marginally affected, and this scenario may be applicable both to fly-by vortices and double reconnections. Also Kelvin modes can be excited in the collision \cite{Leadbeater01,Vinen05,Fonda14} but, if present, they seem not to affect the lifetime, while they are likely responsible for the change of visibility of the vortices, as they can produce out-of-plane distortions and hence a change of contrast in the density distribution. Finally, the observation of a shorter lifetime in
configurations with $N_V(0)=3$ can be understood by considering the role of a third vortex in the collision of two other vortices, whose tendency to rotate in the radial plane is frustrated by three-body interaction, thus enhancing the probability of collisions and reconnections. A similar role of three-body interactions in the dynamics of vortices was recently investigated in the context of 2D classical turbulence \cite{Sire11}.

Our experimental results demand new theoretical models. So far, numerical simulations of vortex reconnection are usually performed with vortex lines initially at rest, at small distance, which then evolve in time \cite{Koplik93,Nazarenko03,Gabbay98,Zuccher12,Wells15}, while in our case the role of the relative velocity seems to be crucial. Shedding light on this, and generally on the dynamics of few vortices in such a relatively simple configuration, can help to understand the physics of vorticity in more complex settings, like those of Refs.~\cite{Henn09,Seman10,Yukalov15}, in the search of a satisfactory comprehension of quantum turbulence in superfluids with boundaries. 

\begin{acknowledgments}
We thank L.P. Pitaevskii, N.P. Proukakis, I-Kang Liu, N.G. Parker and C.F. Barenghi for insightful discussions. We acknowledge Provincia Autonoma di Trento for funding.   
\end{acknowledgments}

%\bibliography{SVDynamics}

\begin{thebibliography}{63}%
\makeatletter
\providecommand \@ifxundefined [1]{%
 \@ifx{#1\undefined}
}%
\providecommand \@ifnum [1]{%
 \ifnum #1\expandafter \@firstoftwo
 \else \expandafter \@secondoftwo
 \fi
}%
\providecommand \@ifx [1]{%
 \ifx #1\expandafter \@firstoftwo
 \else \expandafter \@secondoftwo
 \fi
}%
\providecommand \natexlab [1]{#1}%
\providecommand \enquote  [1]{``#1''}%
\providecommand \bibnamefont  [1]{#1}%
\providecommand \bibfnamefont [1]{#1}%
\providecommand \citenamefont [1]{#1}%
\providecommand \href@noop [0]{\@secondoftwo}%
\providecommand \href [0]{\begingroup \@sanitize@url \@href}%
\providecommand \@href[1]{\@@startlink{#1}\@@href}%
\providecommand \@@href[1]{\endgroup#1\@@endlink}%
\providecommand \@sanitize@url [0]{\catcode `\\12\catcode `\$12\catcode
  `\&12\catcode `\#12\catcode `\^12\catcode `\_12\catcode `\%12\relax}%
\providecommand \@@startlink[1]{}%
\providecommand \@@endlink[0]{}%
\providecommand \url  [0]{\begingroup\@sanitize@url \@url }%
\providecommand \@url [1]{\endgroup\@href {#1}{\urlprefix }}%
\providecommand \urlprefix  [0]{URL }%
\providecommand \Eprint [0]{\href }%
\providecommand \doibase [0]{http://dx.doi.org/}%
\providecommand \selectlanguage [0]{\@gobble}%
\providecommand \bibinfo  [0]{\@secondoftwo}%
\providecommand \bibfield  [0]{\@secondoftwo}%
\providecommand \translation [1]{[#1]}%
\providecommand \BibitemOpen [0]{}%
\providecommand \bibitemStop [0]{}%
\providecommand \bibitemNoStop [0]{.\EOS\space}%
\providecommand \EOS [0]{\spacefactor3000\relax}%
\providecommand \BibitemShut  [1]{\csname bibitem#1\endcsname}%
\let\auto@bib@innerbib\@empty
%</preamble>
\bibitem [{\citenamefont {Feynman}(1955)}]{Feynman55}%
  \BibitemOpen
  \bibfield  {author} {\bibinfo {author} {\bibfnamefont {R.~P.}\ \bibnamefont
  {Feynman}},\ }in\ \href {\doibase 10.1016/S0079-6417(08)60077-3} {\emph
  {\bibinfo {booktitle} {Progress in Low Temperature Physics}}},\ Vol.~\bibinfo
  {volume} {1},\ \bibinfo {editor} {edited by\ \bibinfo {editor} {\bibfnamefont
  {C.}~\bibnamefont {Gorter}}}\ (\bibinfo  {publisher} {Elsevier},\ \bibinfo
  {year} {1955})\ p.~\bibinfo {pages} {17}\BibitemShut {NoStop}%
\bibitem [{\citenamefont {Donnelly}(1991)}]{Donnelly91}%
  \BibitemOpen
  \bibfield  {author} {\bibinfo {author} {\bibfnamefont {R.~J.}\ \bibnamefont
  {Donnelly}},\ }\href@noop {} {\emph {\bibinfo {title} {Quantized Vortices in
  Helium II}}}\ (\bibinfo  {publisher} {Cambridge University Press},\ \bibinfo
  {year} {1991})\ p.\ \bibinfo {pages} {346}\BibitemShut {NoStop}%
\bibitem [{\citenamefont {Tinkham}(1996)}]{Tinkham96}%
  \BibitemOpen
  \bibfield  {author} {\bibinfo {author} {\bibfnamefont {M.}~\bibnamefont
  {Tinkham}},\ }\href@noop {} {\emph {\bibinfo {title} {Introduction to
  superconductivity}}}\ (\bibinfo  {publisher} {Dover Publications},\ \bibinfo
  {year} {1996})\ p.\ \bibinfo {pages} {480}\BibitemShut {NoStop}%
\bibitem [{\citenamefont {Baym}\ and\ \citenamefont {Pethick}(1975)}]{Baym75}%
  \BibitemOpen
  \bibfield  {author} {\bibinfo {author} {\bibfnamefont {G.}~\bibnamefont
  {Baym}}\ and\ \bibinfo {author} {\bibfnamefont {C.}~\bibnamefont {Pethick}},\
  }\href {\doibase 10.1146/annurev.ns.25.120175.000331} {\bibfield  {journal}
  {\bibinfo  {journal} {Ann. Rev. Nucl. Sci.}\ }\textbf {\bibinfo {volume}
  {25}},\ \bibinfo {pages} {27} (\bibinfo {year} {1975})}\BibitemShut {NoStop}%
\bibitem [{\citenamefont {Lalescu}\ \emph {et~al.}(2015)\citenamefont
  {Lalescu}, \citenamefont {Shi}, \citenamefont {Eyink}, \citenamefont
  {Drivas}, \citenamefont {Vishniac},\ and\ \citenamefont
  {Lazarian}}]{Lalescu15}%
  \BibitemOpen
  \bibfield  {author} {\bibinfo {author} {\bibfnamefont {C.~C.}\ \bibnamefont
  {Lalescu}}, \bibinfo {author} {\bibfnamefont {Y.-K.}\ \bibnamefont {Shi}},
  \bibinfo {author} {\bibfnamefont {G.~L.}\ \bibnamefont {Eyink}}, \bibinfo
  {author} {\bibfnamefont {T.~D.}\ \bibnamefont {Drivas}}, \bibinfo {author}
  {\bibfnamefont {E.~T.}\ \bibnamefont {Vishniac}}, \ and\ \bibinfo {author}
  {\bibfnamefont {A.}~\bibnamefont {Lazarian}},\ }\href {\doibase
  10.1103/PhysRevLett.115.025001} {\bibfield  {journal} {\bibinfo  {journal}
  {Phys. Rev. Lett.}\ }\textbf {\bibinfo {volume} {115}},\ \bibinfo {pages}
  {025001} (\bibinfo {year} {2015})}\BibitemShut {NoStop}%
\bibitem [{\citenamefont {Paoletti}\ and\ \citenamefont
  {Lathrop}(2011)}]{Paoletti11}%
  \BibitemOpen
  \bibfield  {author} {\bibinfo {author} {\bibfnamefont {M.~S.}\ \bibnamefont
  {Paoletti}}\ and\ \bibinfo {author} {\bibfnamefont {D.~P.}\ \bibnamefont
  {Lathrop}},\ }\href {\doibase 10.1146/annurev-conmatphys-062910-140533}
  {\bibfield  {journal} {\bibinfo  {journal} {Ann. Rev. Cond. Matter Phys.}\
  }\textbf {\bibinfo {volume} {2}},\ \bibinfo {pages} {213} (\bibinfo {year}
  {2011})}\BibitemShut {NoStop}%
\bibitem [{\citenamefont {Zuccher}\ \emph {et~al.}(2012)\citenamefont
  {Zuccher}, \citenamefont {Caliari}, \citenamefont {Baggaley},\ and\
  \citenamefont {Barenghi}}]{Zuccher12}%
  \BibitemOpen
  \bibfield  {author} {\bibinfo {author} {\bibfnamefont {S.}~\bibnamefont
  {Zuccher}}, \bibinfo {author} {\bibfnamefont {M.}~\bibnamefont {Caliari}},
  \bibinfo {author} {\bibfnamefont {A.}~\bibnamefont {Baggaley}}, \ and\
  \bibinfo {author} {\bibfnamefont {C.}~\bibnamefont {Barenghi}},\ }\href
  {\doibase 10.1063/1.4772198} {\bibfield  {journal} {\bibinfo  {journal}
  {Phys. of Fluids}\ }\textbf {\bibinfo {volume} {24}},\ \bibinfo {pages}
  {125108} (\bibinfo {year} {2012})}\BibitemShut {NoStop}%
\bibitem [{\citenamefont {Fetter}(2009)}]{Fetter09}%
  \BibitemOpen
  \bibfield  {author} {\bibinfo {author} {\bibfnamefont {A.}~\bibnamefont
  {Fetter}},\ }\href {\doibase 10.1103/RevModPhys.81,647} {\bibfield  {journal}
  {\bibinfo  {journal} {Rev. Mod. Phys.}\ }\textbf {\bibinfo {volume} {81}},\
  \bibinfo {pages} {647} (\bibinfo {year} {2009})}\BibitemShut {NoStop}%
\bibitem [{\citenamefont {Madison}\ \emph {et~al.}(2000)\citenamefont
  {Madison}, \citenamefont {Chevy}, \citenamefont {Wohlleben},\ and\
  \citenamefont {Dalibard}}]{Madison00}%
  \BibitemOpen
  \bibfield  {author} {\bibinfo {author} {\bibfnamefont {K.~W.}\ \bibnamefont
  {Madison}}, \bibinfo {author} {\bibfnamefont {F.}~\bibnamefont {Chevy}},
  \bibinfo {author} {\bibfnamefont {W.}~\bibnamefont {Wohlleben}}, \ and\
  \bibinfo {author} {\bibfnamefont {J.}~\bibnamefont {Dalibard}},\ }\href
  {\doibase 10.1103/PhysRevLett.84.806} {\bibfield  {journal} {\bibinfo
  {journal} {Phys. Rev. Lett.}\ }\textbf {\bibinfo {volume} {84}},\ \bibinfo
  {pages} {806} (\bibinfo {year} {2000})}\BibitemShut {NoStop}%
\bibitem [{\citenamefont {Abo-Shaeer}\ \emph {et~al.}(2001)\citenamefont
  {Abo-Shaeer}, \citenamefont {Raman}, \citenamefont {Vogels},\ and\
  \citenamefont {Ketterle}}]{AboShaeer01}%
  \BibitemOpen
  \bibfield  {author} {\bibinfo {author} {\bibfnamefont {J.}~\bibnamefont
  {Abo-Shaeer}}, \bibinfo {author} {\bibfnamefont {C.}~\bibnamefont {Raman}},
  \bibinfo {author} {\bibfnamefont {J.}~\bibnamefont {Vogels}}, \ and\ \bibinfo
  {author} {\bibfnamefont {W.}~\bibnamefont {Ketterle}},\ }\href@noop {}
  {\bibfield  {journal} {\bibinfo  {journal} {Science}\ }\textbf {\bibinfo
  {volume} {292}},\ \bibinfo {pages} {476} (\bibinfo {year}
  {2001})}\BibitemShut {NoStop}%
\bibitem [{\citenamefont {Engels}\ \emph {et~al.}(2002)\citenamefont {Engels},
  \citenamefont {Coddington}, \citenamefont {Haljan},\ and\ \citenamefont
  {Cornell}}]{Engels02}%
  \BibitemOpen
  \bibfield  {author} {\bibinfo {author} {\bibfnamefont {P.}~\bibnamefont
  {Engels}}, \bibinfo {author} {\bibfnamefont {I.}~\bibnamefont {Coddington}},
  \bibinfo {author} {\bibfnamefont {P.~C.}\ \bibnamefont {Haljan}}, \ and\
  \bibinfo {author} {\bibfnamefont {E.~A.}\ \bibnamefont {Cornell}},\ }\href
  {\doibase 10.1103/PhysRevLett.89.100403} {\bibfield  {journal} {\bibinfo
  {journal} {Phys. Rev. Lett.}\ }\textbf {\bibinfo {volume} {89}},\ \bibinfo
  {pages} {100403} (\bibinfo {year} {2002})}\BibitemShut {NoStop}%
\bibitem [{\citenamefont {Engels}\ \emph {et~al.}(2003)\citenamefont {Engels},
  \citenamefont {Coddington}, \citenamefont {Haljan}, \citenamefont
  {Schweikhard},\ and\ \citenamefont {Cornell}}]{Engels03}%
  \BibitemOpen
  \bibfield  {author} {\bibinfo {author} {\bibfnamefont {P.}~\bibnamefont
  {Engels}}, \bibinfo {author} {\bibfnamefont {I.}~\bibnamefont {Coddington}},
  \bibinfo {author} {\bibfnamefont {P.~C.}\ \bibnamefont {Haljan}}, \bibinfo
  {author} {\bibfnamefont {V.}~\bibnamefont {Schweikhard}}, \ and\ \bibinfo
  {author} {\bibfnamefont {E.~A.}\ \bibnamefont {Cornell}},\ }\href {\doibase
  10.1103/PhysRevLett.90.170405} {\bibfield  {journal} {\bibinfo  {journal}
  {Phys. Rev. Lett.}\ }\textbf {\bibinfo {volume} {90}},\ \bibinfo {pages}
  {170405} (\bibinfo {year} {2003})}\BibitemShut {NoStop}%
\bibitem [{\citenamefont {Coddington}\ \emph {et~al.}(2003)\citenamefont
  {Coddington}, \citenamefont {Engels}, \citenamefont {Schweikhard},\ and\
  \citenamefont {Cornell}}]{Coddington03}%
  \BibitemOpen
  \bibfield  {author} {\bibinfo {author} {\bibfnamefont {I.}~\bibnamefont
  {Coddington}}, \bibinfo {author} {\bibfnamefont {P.}~\bibnamefont {Engels}},
  \bibinfo {author} {\bibfnamefont {V.}~\bibnamefont {Schweikhard}}, \ and\
  \bibinfo {author} {\bibfnamefont {E.~A.}\ \bibnamefont {Cornell}},\ }\href
  {\doibase 10.1103/PhysRevLett.91.100402} {\bibfield  {journal} {\bibinfo
  {journal} {Phys. Rev. Lett.}\ }\textbf {\bibinfo {volume} {91}},\ \bibinfo
  {pages} {100402} (\bibinfo {year} {2003})}\BibitemShut {NoStop}%
\bibitem [{\citenamefont {Weiler}\ \emph {et~al.}(2008)\citenamefont {Weiler},
  \citenamefont {Neely}, \citenamefont {Scherer}, \citenamefont {Bradley},
  \citenamefont {Davis},\ and\ \citenamefont {Anderson}}]{Weiler08}%
  \BibitemOpen
  \bibfield  {author} {\bibinfo {author} {\bibfnamefont {C.~N.}\ \bibnamefont
  {Weiler}}, \bibinfo {author} {\bibfnamefont {T.~W.}\ \bibnamefont {Neely}},
  \bibinfo {author} {\bibfnamefont {D.~R.}\ \bibnamefont {Scherer}}, \bibinfo
  {author} {\bibfnamefont {A.~S.}\ \bibnamefont {Bradley}}, \bibinfo {author}
  {\bibfnamefont {M.~J.}\ \bibnamefont {Davis}}, \ and\ \bibinfo {author}
  {\bibfnamefont {B.~P.}\ \bibnamefont {Anderson}},\ }\href {\doibase
  10.1038/nature07334} {\bibfield  {journal} {\bibinfo  {journal} {Nature}\
  }\textbf {\bibinfo {volume} {455}},\ \bibinfo {pages} {948} (\bibinfo {year}
  {2008})}\BibitemShut {NoStop}%
\bibitem [{\citenamefont {Neely}\ \emph {et~al.}(2010)\citenamefont {Neely},
  \citenamefont {Samson}, \citenamefont {Bradley}, \citenamefont {Davis},\ and\
  \citenamefont {Anderson}}]{Neely10}%
  \BibitemOpen
  \bibfield  {author} {\bibinfo {author} {\bibfnamefont {T.~W.}\ \bibnamefont
  {Neely}}, \bibinfo {author} {\bibfnamefont {E.~C.}\ \bibnamefont {Samson}},
  \bibinfo {author} {\bibfnamefont {A.~S.}\ \bibnamefont {Bradley}}, \bibinfo
  {author} {\bibfnamefont {M.~J.}\ \bibnamefont {Davis}}, \ and\ \bibinfo
  {author} {\bibfnamefont {B.~P.}\ \bibnamefont {Anderson}},\ }\href {\doibase
  10.1103/PhysRevLett.104.160401} {\bibfield  {journal} {\bibinfo  {journal}
  {Phys. Rev. Lett.}\ }\textbf {\bibinfo {volume} {104}},\ \bibinfo {pages}
  {160401} (\bibinfo {year} {2010})}\BibitemShut {NoStop}%
\bibitem [{\citenamefont {Freilich}\ \emph {et~al.}(2010)\citenamefont
  {Freilich}, \citenamefont {Bianchi}, \citenamefont {Kaufman}, \citenamefont
  {Langin},\ and\ \citenamefont {Hall}}]{Freilich10}%
  \BibitemOpen
  \bibfield  {author} {\bibinfo {author} {\bibfnamefont {D.~V.}\ \bibnamefont
  {Freilich}}, \bibinfo {author} {\bibfnamefont {D.~M.}\ \bibnamefont
  {Bianchi}}, \bibinfo {author} {\bibfnamefont {A.~M.}\ \bibnamefont
  {Kaufman}}, \bibinfo {author} {\bibfnamefont {T.~K.}\ \bibnamefont {Langin}},
  \ and\ \bibinfo {author} {\bibfnamefont {D.~S.}\ \bibnamefont {Hall}},\
  }\href@noop {} {\bibfield  {journal} {\bibinfo  {journal} {Science}\ }\textbf
  {\bibinfo {volume} {329}},\ \bibinfo {pages} {1182} (\bibinfo {year}
  {2010})}\BibitemShut {NoStop}%
\bibitem [{\citenamefont {Middelkamp}\ \emph {et~al.}(2011)\citenamefont
  {Middelkamp}, \citenamefont {Torres}, \citenamefont {Kevrekidis},
  \citenamefont {Frantzeskakis}, \citenamefont {Carretero-Gonz{\'a}lez},
  \citenamefont {Schmelcher}, \citenamefont {Freilich},\ and\ \citenamefont
  {Hall}}]{Middelkamp11}%
  \BibitemOpen
  \bibfield  {author} {\bibinfo {author} {\bibfnamefont {S.}~\bibnamefont
  {Middelkamp}}, \bibinfo {author} {\bibfnamefont {P.}~\bibnamefont {Torres}},
  \bibinfo {author} {\bibfnamefont {P.~G.}\ \bibnamefont {Kevrekidis}},
  \bibinfo {author} {\bibfnamefont {D.~J.}\ \bibnamefont {Frantzeskakis}},
  \bibinfo {author} {\bibfnamefont {R.}~\bibnamefont {Carretero-Gonz{\'a}lez}},
  \bibinfo {author} {\bibfnamefont {P.}~\bibnamefont {Schmelcher}}, \bibinfo
  {author} {\bibfnamefont {D.~V.}\ \bibnamefont {Freilich}}, \ and\ \bibinfo
  {author} {\bibfnamefont {D.~S.}\ \bibnamefont {Hall}},\ }\href {\doibase
  10.1103/PhysRevA.84.011605} {\bibfield  {journal} {\bibinfo  {journal} {Phys.
  Rev. A}\ }\textbf {\bibinfo {volume} {84}},\ \bibinfo {pages} {011605(R)}
  (\bibinfo {year} {2011})}\BibitemShut {NoStop}%
\bibitem [{\citenamefont {Navarro}\ \emph {et~al.}(2013)\citenamefont
  {Navarro}, \citenamefont {Carretero-Gonz\'alez}, \citenamefont {Torres},
  \citenamefont {Kevrekidis}, \citenamefont {Frantzeskakis}, \citenamefont
  {Ray}, \citenamefont {Altunta\ifmmode~\mbox{\c{s}}\else \c{s}\fi{}},\ and\
  \citenamefont {Hall}}]{Navarro13}%
  \BibitemOpen
  \bibfield  {author} {\bibinfo {author} {\bibfnamefont {R.}~\bibnamefont
  {Navarro}}, \bibinfo {author} {\bibfnamefont {R.}~\bibnamefont
  {Carretero-Gonz\'alez}}, \bibinfo {author} {\bibfnamefont {P.~J.}\
  \bibnamefont {Torres}}, \bibinfo {author} {\bibfnamefont {P.~G.}\
  \bibnamefont {Kevrekidis}}, \bibinfo {author} {\bibfnamefont {D.~J.}\
  \bibnamefont {Frantzeskakis}}, \bibinfo {author} {\bibfnamefont {M.~W.}\
  \bibnamefont {Ray}}, \bibinfo {author} {\bibfnamefont {E.}~\bibnamefont
  {Altunta\ifmmode~\mbox{\c{s}}\else \c{s}\fi{}}}, \ and\ \bibinfo {author}
  {\bibfnamefont {D.~S.}\ \bibnamefont {Hall}},\ }\href {\doibase
  10.1103/PhysRevLett.110.225301} {\bibfield  {journal} {\bibinfo  {journal}
  {Phys. Rev. Lett.}\ }\textbf {\bibinfo {volume} {110}},\ \bibinfo {pages}
  {225301} (\bibinfo {year} {2013})}\BibitemShut {NoStop}%
\bibitem [{\citenamefont {Kwon}\ \emph {et~al.}(2014)\citenamefont {Kwon},
  \citenamefont {Moon}, \citenamefont {Choi}, \citenamefont {Seo},\ and\
  \citenamefont {Shin}}]{Kwon14}%
  \BibitemOpen
  \bibfield  {author} {\bibinfo {author} {\bibfnamefont {W.~J.}\ \bibnamefont
  {Kwon}}, \bibinfo {author} {\bibfnamefont {G.}~\bibnamefont {Moon}}, \bibinfo
  {author} {\bibfnamefont {J.-y.}\ \bibnamefont {Choi}}, \bibinfo {author}
  {\bibfnamefont {S.~W.}\ \bibnamefont {Seo}}, \ and\ \bibinfo {author}
  {\bibfnamefont {Y.-i.}\ \bibnamefont {Shin}},\ }\href {\doibase
  10.1103/PhysRevA.90.063627} {\bibfield  {journal} {\bibinfo  {journal} {Phys.
  Rev. A}\ }\textbf {\bibinfo {volume} {90}},\ \bibinfo {pages} {063627}
  (\bibinfo {year} {2014})}\BibitemShut {NoStop}%
\bibitem [{\citenamefont {Brand}\ and\ \citenamefont
  {Reinhardt}(2002)}]{Brand02}%
  \BibitemOpen
  \bibfield  {author} {\bibinfo {author} {\bibfnamefont {J.}~\bibnamefont
  {Brand}}\ and\ \bibinfo {author} {\bibfnamefont {W.~P.}\ \bibnamefont
  {Reinhardt}},\ }\href {\doibase 10.1103/PhysRevA.65.043612} {\bibfield
  {journal} {\bibinfo  {journal} {Phys. Rev. A}\ }\textbf {\bibinfo {volume}
  {65}},\ \bibinfo {pages} {043612} (\bibinfo {year} {2002})}\BibitemShut
  {NoStop}%
\bibitem [{\citenamefont {Komineas}\ and\ \citenamefont
  {Papanicolaou}(2003)}]{Komineas03}%
  \BibitemOpen
  \bibfield  {author} {\bibinfo {author} {\bibfnamefont {S.}~\bibnamefont
  {Komineas}}\ and\ \bibinfo {author} {\bibfnamefont {N.}~\bibnamefont
  {Papanicolaou}},\ }\href {\doibase 10.1103/PhysRevA.68.043617} {\bibfield
  {journal} {\bibinfo  {journal} {Phys. Rev. A}\ }\textbf {\bibinfo {volume}
  {68}},\ \bibinfo {pages} {043617} (\bibinfo {year} {2003})}\BibitemShut
  {NoStop}%
\bibitem [{\citenamefont {Donadello}\ \emph {et~al.}(2014)\citenamefont
  {Donadello}, \citenamefont {Serafini}, \citenamefont {Tylutki}, \citenamefont
  {Pitaevskii}, \citenamefont {Dalfovo}, \citenamefont {Lamporesi},\ and\
  \citenamefont {Ferrari}}]{Donadello14}%
  \BibitemOpen
  \bibfield  {author} {\bibinfo {author} {\bibfnamefont {S.}~\bibnamefont
  {Donadello}}, \bibinfo {author} {\bibfnamefont {S.}~\bibnamefont {Serafini}},
  \bibinfo {author} {\bibfnamefont {M.}~\bibnamefont {Tylutki}}, \bibinfo
  {author} {\bibfnamefont {L.~P.}\ \bibnamefont {Pitaevskii}}, \bibinfo
  {author} {\bibfnamefont {F.}~\bibnamefont {Dalfovo}}, \bibinfo {author}
  {\bibfnamefont {G.}~\bibnamefont {Lamporesi}}, \ and\ \bibinfo {author}
  {\bibfnamefont {G.}~\bibnamefont {Ferrari}},\ }\href {\doibase
  10.1103/PhysRevLett.113.065302} {\bibfield  {journal} {\bibinfo  {journal}
  {Phys. Rev. Lett.}\ }\textbf {\bibinfo {volume} {113}},\ \bibinfo {pages}
  {065302} (\bibinfo {year} {2014})}\BibitemShut {NoStop}%
\bibitem [{\citenamefont {Tylutki}\ \emph {et~al.}(2015)\citenamefont
  {Tylutki}, \citenamefont {Donadello}, \citenamefont {Serafini}, \citenamefont
  {Pitaevskii}, \citenamefont {Dalfovo}, \citenamefont {Lamporesi},\ and\
  \citenamefont {Ferrari}}]{Tylutki15}%
  \BibitemOpen
  \bibfield  {author} {\bibinfo {author} {\bibfnamefont {M.}~\bibnamefont
  {Tylutki}}, \bibinfo {author} {\bibfnamefont {S.}~\bibnamefont {Donadello}},
  \bibinfo {author} {\bibfnamefont {S.}~\bibnamefont {Serafini}}, \bibinfo
  {author} {\bibfnamefont {L.~P.}\ \bibnamefont {Pitaevskii}}, \bibinfo
  {author} {\bibfnamefont {F.}~\bibnamefont {Dalfovo}}, \bibinfo {author}
  {\bibfnamefont {G.}~\bibnamefont {Lamporesi}}, \ and\ \bibinfo {author}
  {\bibfnamefont {G.}~\bibnamefont {Ferrari}},\ }\href {\doibase
  10.1140/epjst/e2015-02389-7} {\bibfield  {journal} {\bibinfo  {journal} {Eur.
  Phys. J. Special Topics}\ }\textbf {\bibinfo {volume} {224}},\ \bibinfo
  {pages} {577} (\bibinfo {year} {2015})}\BibitemShut {NoStop}%
\bibitem [{\citenamefont {Ku}\ \emph {et~al.}(2014)\citenamefont {Ku},
  \citenamefont {Ji}, \citenamefont {Mukherjee}, \citenamefont
  {Guardado-Sanchez}, \citenamefont {Cheuk}, \citenamefont {Yefsah},\ and\
  \citenamefont {Zwierlein}}]{Ku14}%
  \BibitemOpen
  \bibfield  {author} {\bibinfo {author} {\bibfnamefont {M.~J.~H.}\
  \bibnamefont {Ku}}, \bibinfo {author} {\bibfnamefont {W.}~\bibnamefont {Ji}},
  \bibinfo {author} {\bibfnamefont {B.}~\bibnamefont {Mukherjee}}, \bibinfo
  {author} {\bibfnamefont {E.}~\bibnamefont {Guardado-Sanchez}}, \bibinfo
  {author} {\bibfnamefont {L.~W.}\ \bibnamefont {Cheuk}}, \bibinfo {author}
  {\bibfnamefont {T.}~\bibnamefont {Yefsah}}, \ and\ \bibinfo {author}
  {\bibfnamefont {M.~W.}\ \bibnamefont {Zwierlein}},\ }\href@noop {} {\bibfield
   {journal} {\bibinfo  {journal} {Phys. Rev. Lett.}\ }\textbf {\bibinfo
  {volume} {113}},\ \bibinfo {pages} {065301} (\bibinfo {year}
  {2014})}\BibitemShut {NoStop}%
\bibitem [{\citenamefont {Lamporesi}\ \emph
  {et~al.}(2013{\natexlab{a}})\citenamefont {Lamporesi}, \citenamefont
  {Donadello}, \citenamefont {Serafini},\ and\ \citenamefont
  {Ferrari}}]{Lamporesi13}%
  \BibitemOpen
  \bibfield  {author} {\bibinfo {author} {\bibfnamefont {G.}~\bibnamefont
  {Lamporesi}}, \bibinfo {author} {\bibfnamefont {S.}~\bibnamefont
  {Donadello}}, \bibinfo {author} {\bibfnamefont {S.}~\bibnamefont {Serafini}},
  \ and\ \bibinfo {author} {\bibfnamefont {G.}~\bibnamefont {Ferrari}},\
  }\href@noop {} {\bibfield  {journal} {\bibinfo  {journal} {Rev. Sci.
  Instrum.}\ }\textbf {\bibinfo {volume} {84}},\ \bibinfo {pages} {063102}
  (\bibinfo {year} {2013}{\natexlab{a}})}\BibitemShut {NoStop}%
\bibitem [{\citenamefont {Kibble}(1980)}]{Kibble80}%
  \BibitemOpen
  \bibfield  {author} {\bibinfo {author} {\bibfnamefont {T.}~\bibnamefont
  {Kibble}},\ }\href {\doibase 10.1016/0370-1573(80)90091-5} {\bibfield
  {journal} {\bibinfo  {journal} {Phys. Rep.}\ }\textbf {\bibinfo {volume}
  {67}},\ \bibinfo {pages} {183 } (\bibinfo {year} {1980})}\BibitemShut
  {NoStop}%
\bibitem [{\citenamefont {Zurek}(1985)}]{Zurek85}%
  \BibitemOpen
  \bibfield  {author} {\bibinfo {author} {\bibfnamefont {W.~H.}\ \bibnamefont
  {Zurek}},\ }\href@noop {} {\bibfield  {journal} {\bibinfo  {journal}
  {Nature}\ }\textbf {\bibinfo {volume} {317}},\ \bibinfo {pages} {505}
  (\bibinfo {year} {1985})}\BibitemShut {NoStop}%
\bibitem [{\citenamefont {Lamporesi}\ \emph
  {et~al.}(2013{\natexlab{b}})\citenamefont {Lamporesi}, \citenamefont
  {Donadello}, \citenamefont {Serafini}, \citenamefont {Dalfovo},\ and\
  \citenamefont {Ferrari}}]{LamporesiKZM13}%
  \BibitemOpen
  \bibfield  {author} {\bibinfo {author} {\bibfnamefont {G.}~\bibnamefont
  {Lamporesi}}, \bibinfo {author} {\bibfnamefont {S.}~\bibnamefont
  {Donadello}}, \bibinfo {author} {\bibfnamefont {S.}~\bibnamefont {Serafini}},
  \bibinfo {author} {\bibfnamefont {F.}~\bibnamefont {Dalfovo}}, \ and\
  \bibinfo {author} {\bibfnamefont {G.}~\bibnamefont {Ferrari}},\ }\href@noop
  {} {\bibfield  {journal} {\bibinfo  {journal} {Nat. Phys.}\ }\textbf
  {\bibinfo {volume} {9}},\ \bibinfo {pages} {656} (\bibinfo {year}
  {2013}{\natexlab{b}})}\BibitemShut {NoStop}%
\bibitem [{\citenamefont {Ramanathan}\ \emph {et~al.}(2012)\citenamefont
  {Ramanathan}, \citenamefont {Muniz}, \citenamefont {Wright}, \citenamefont
  {Anderson}, \citenamefont {Phillips}, \citenamefont {Helmerson},\ and\
  \citenamefont {Campbell}}]{Ramanathan12}%
  \BibitemOpen
  \bibfield  {author} {\bibinfo {author} {\bibfnamefont {A.}~\bibnamefont
  {Ramanathan}}, \bibinfo {author} {\bibfnamefont {S.~R.}\ \bibnamefont
  {Muniz}}, \bibinfo {author} {\bibfnamefont {K.~C.}\ \bibnamefont {Wright}},
  \bibinfo {author} {\bibfnamefont {R.~P.}\ \bibnamefont {Anderson}}, \bibinfo
  {author} {\bibfnamefont {W.~D.}\ \bibnamefont {Phillips}}, \bibinfo {author}
  {\bibfnamefont {K.}~\bibnamefont {Helmerson}}, \ and\ \bibinfo {author}
  {\bibfnamefont {G.~K.}\ \bibnamefont {Campbell}},\ }\href {\doibase
  http://dx.doi.org/10.1063/1.4747163} {\bibfield  {journal} {\bibinfo
  {journal} {Review of Scientific Instruments}\ }\textbf {\bibinfo {volume}
  {83}},\ \bibinfo {eid} {083119} (\bibinfo {year} {2012})}\BibitemShut
  {NoStop}%
\bibitem [{SM()}]{SM}%
  \BibitemOpen
  \href@noop {} {\bibinfo {title} {\rm See Supplemental Material, which includes Ref. \cite{Zobay01} for RF dressing of hyperfine levels}\
  }\BibitemShut {NoStop}%
\bibitem[{\citenamefont{Zobay and Garraway}(2001)}]{Zobay01}
\bibinfo{author}{\bibfnamefont{O.}~\bibnamefont{Zobay}} \bibnamefont{and}
  \bibinfo{author}{\bibfnamefont{B.~M.} \bibnamefont{Garraway}},
  \bibinfo{journal}{Phys. Rev. Lett.} \textbf{\bibinfo{volume}{86}},
  \bibinfo{pages}{1195} (\bibinfo{year}{2001})\BibitemShut
  {NoStop}%
\bibitem [{\citenamefont {Dalfovo}\ \emph {et~al.}(1999)\citenamefont
  {Dalfovo}, \citenamefont {Giorgini}, \citenamefont {Pitaevskii},\ and\
  \citenamefont {Stringari}}]{Dalfovo99}%
  \BibitemOpen
  \bibfield  {author} {\bibinfo {author} {\bibfnamefont {F.}~\bibnamefont
  {Dalfovo}}, \bibinfo {author} {\bibfnamefont {S.}~\bibnamefont {Giorgini}},
  \bibinfo {author} {\bibfnamefont {L.~P.}\ \bibnamefont {Pitaevskii}}, \ and\
  \bibinfo {author} {\bibfnamefont {S.}~\bibnamefont {Stringari}},\ }\href
  {\doibase 10.1103/RevModPhys.71.463} {\bibfield  {journal} {\bibinfo
  {journal} {Rev. Mod. Phys.}\ }\textbf {\bibinfo {volume} {71}},\ \bibinfo
  {pages} {463} (\bibinfo {year} {1999})}\BibitemShut {NoStop}%
\bibitem [{\citenamefont {Anderson}\ \emph {et~al.}(2000)\citenamefont
  {Anderson}, \citenamefont {Haljan}, \citenamefont {Wieman},\ and\
  \citenamefont {Cornell}}]{Anderson2000}%
  \BibitemOpen
  \bibfield  {author} {\bibinfo {author} {\bibfnamefont {B.~P.}\ \bibnamefont
  {Anderson}}, \bibinfo {author} {\bibfnamefont {P.~C.}\ \bibnamefont
  {Haljan}}, \bibinfo {author} {\bibfnamefont {C.~E.}\ \bibnamefont {Wieman}},
  \ and\ \bibinfo {author} {\bibfnamefont {E.~A.}\ \bibnamefont {Cornell}},\
  }\href {\doibase 10.1103/PhysRevLett.85.2857} {\bibfield  {journal} {\bibinfo
   {journal} {Phys. Rev. Lett.}\ }\textbf {\bibinfo {volume} {85}},\ \bibinfo
  {pages} {2857} (\bibinfo {year} {2000})}\BibitemShut {NoStop}%
\bibitem [{\citenamefont {Sheehy}\ and\ \citenamefont
  {Radzihovsky}(2004)}]{Sheehy04}%
  \BibitemOpen
  \bibfield  {author} {\bibinfo {author} {\bibfnamefont {D.}~\bibnamefont
  {Sheehy}}\ and\ \bibinfo {author} {\bibfnamefont {L.}~\bibnamefont
  {Radzihovsky}},\ }\href {\doibase 10.1103/PhysRevA.70.063620} {\bibfield
  {journal} {\bibinfo  {journal} {Phys. Rev. A}\ }\textbf {\bibinfo {volume}
  {70}},\ \bibinfo {pages} {063620} (\bibinfo {year} {2004})}\BibitemShut
  {NoStop}%
\bibitem [{\citenamefont {Svidzinsky}\ and\ \citenamefont
  {Fetter}(2000{\natexlab{a}})}]{Svidzinsky00}%
  \BibitemOpen
  \bibfield  {author} {\bibinfo {author} {\bibfnamefont {A.}~\bibnamefont
  {Svidzinsky}}\ and\ \bibinfo {author} {\bibfnamefont {A.}~\bibnamefont
  {Fetter}},\ }\href@noop {} {\bibfield  {journal} {\bibinfo  {journal} {Phys.
  Rev. Lett.}\ }\textbf {\bibinfo {volume} {84}},\ \bibinfo {pages} {5919}
  (\bibinfo {year} {2000}{\natexlab{a}})}\BibitemShut {NoStop}%
\bibitem [{\citenamefont {Fetter}\ and\ \citenamefont {Kim}(2001)}]{Fetter01}%
  \BibitemOpen
  \bibfield  {author} {\bibinfo {author} {\bibfnamefont {A.~L.}\ \bibnamefont
  {Fetter}}\ and\ \bibinfo {author} {\bibfnamefont {J.-k.}\ \bibnamefont
  {Kim}},\ }\href@noop {} {\bibfield  {journal} {\bibinfo  {journal} {J. Low
  Temp. Phys.}\ }\textbf {\bibinfo {volume} {125}},\ \bibinfo {pages} {239}
  (\bibinfo {year} {2001})}\BibitemShut {NoStop}%
\bibitem [{\citenamefont {Pitaevskii}(2013)}]{Pitaevskii13}%
  \BibitemOpen
  \bibfield  {author} {\bibinfo {author} {\bibfnamefont {L.}~\bibnamefont
  {Pitaevskii}},\ }\href@noop {} {\bibfield  {journal} {\bibinfo  {journal}
  {arXiv:1311.4693v1}\ } (\bibinfo {year} {2013})}\BibitemShut {NoStop}%
\bibitem [{\citenamefont {Busch}\ and\ \citenamefont {Anglin}(2000)}]{Busch00}%
  \BibitemOpen
  \bibfield  {author} {\bibinfo {author} {\bibfnamefont {T.}~\bibnamefont
  {Busch}}\ and\ \bibinfo {author} {\bibfnamefont {J.}~\bibnamefont {Anglin}},\
  }\href {\doibase 10.1103/PhysRevLett.84.2298} {\bibfield  {journal} {\bibinfo
   {journal} {Phys. Rev. Lett.}\ }\textbf {\bibinfo {volume} {84}},\ \bibinfo
  {pages} {2298} (\bibinfo {year} {2000})}\BibitemShut {NoStop}%
\bibitem [{\citenamefont {Konotop}\ and\ \citenamefont
  {Pitaevskii}(2004)}]{Konotop04}%
  \BibitemOpen
  \bibfield  {author} {\bibinfo {author} {\bibfnamefont {V.~V.}\ \bibnamefont
  {Konotop}}\ and\ \bibinfo {author} {\bibfnamefont {L.}~\bibnamefont
  {Pitaevskii}},\ }\href {\doibase 10.1103/PhysRevLett.93.240403} {\bibfield
  {journal} {\bibinfo  {journal} {Phys. Rev. lett.}\ }\textbf {\bibinfo
  {volume} {93}},\ \bibinfo {pages} {240403} (\bibinfo {year}
  {2004})}\BibitemShut {NoStop}%
\bibitem [{\citenamefont {Svidzinsky}\ and\ \citenamefont
  {Fetter}(2000{\natexlab{b}})}]{Svidzinsky00b}%
  \BibitemOpen
  \bibfield  {author} {\bibinfo {author} {\bibfnamefont {A.}~\bibnamefont
  {Svidzinsky}}\ and\ \bibinfo {author} {\bibfnamefont {A.}~\bibnamefont
  {Fetter}},\ }\href@noop {} {\bibfield  {journal} {\bibinfo  {journal} {Phys.
  Rev. A}\ }\textbf {\bibinfo {volume} {62}},\ \bibinfo {pages} {063617}
  (\bibinfo {year} {2000}{\natexlab{b}})}\BibitemShut {NoStop}%
\bibitem [{\citenamefont {Aftalion}\ and\ \citenamefont
  {Riviere}(2001)}]{Aftalion01}%
  \BibitemOpen
  \bibfield  {author} {\bibinfo {author} {\bibfnamefont {A.}~\bibnamefont
  {Aftalion}}\ and\ \bibinfo {author} {\bibfnamefont {T.}~\bibnamefont
  {Riviere}},\ }\href {\doibase 10.1103/PhysRevA.64.043611} {\bibfield
  {journal} {\bibinfo  {journal} {Phys. Rev. A}\ }\textbf {\bibinfo {volume}
  {64}},\ \bibinfo {pages} {043611} (\bibinfo {year} {2001})}\BibitemShut
  {NoStop}%
\bibitem [{\citenamefont {Garc\'{\i}a-Ripoll}\ and\ \citenamefont
  {P\'erez-Garc\'{\i}a}(2001{\natexlab{a}})}]{Garcia01a}%
  \BibitemOpen
  \bibfield  {author} {\bibinfo {author} {\bibfnamefont {J.}~\bibnamefont
  {Garc\'{\i}a-Ripoll}}\ and\ \bibinfo {author} {\bibfnamefont
  {V.}~\bibnamefont {P\'erez-Garc\'{\i}a}},\ }\href {\doibase
  10.1103/PhysRevA.63.041603} {\bibfield  {journal} {\bibinfo  {journal} {Phys.
  Rev. A}\ }\textbf {\bibinfo {volume} {63}},\ \bibinfo {pages} {041603(R)}
  (\bibinfo {year} {2001}{\natexlab{a}})}\BibitemShut {NoStop}%
\bibitem [{\citenamefont {Garc\'{\i}a-Ripoll}\ and\ \citenamefont
  {P\'erez-Garc\'{\i}a}(2001{\natexlab{b}})}]{Garcia01b}%
  \BibitemOpen
  \bibfield  {author} {\bibinfo {author} {\bibfnamefont {J.}~\bibnamefont
  {Garc\'{\i}a-Ripoll}}\ and\ \bibinfo {author} {\bibfnamefont
  {V.}~\bibnamefont {P\'erez-Garc\'{\i}a}},\ }\href {\doibase
  10.1103/PhysRevA.64.053611} {\bibfield  {journal} {\bibinfo  {journal} {Phys.
  Rev. A}\ }\textbf {\bibinfo {volume} {64}},\ \bibinfo {pages} {053611}
  (\bibinfo {year} {2001}{\natexlab{b}})}\BibitemShut {NoStop}%
\bibitem [{\citenamefont {Modugno}\ \emph {et~al.}(2003)\citenamefont
  {Modugno}, \citenamefont {Pricoupenko},\ and\ \citenamefont
  {Castin}}]{Modugno03}%
  \BibitemOpen
  \bibfield  {author} {\bibinfo {author} {\bibfnamefont {M.}~\bibnamefont
  {Modugno}}, \bibinfo {author} {\bibfnamefont {L.}~\bibnamefont
  {Pricoupenko}}, \ and\ \bibinfo {author} {\bibfnamefont {Y.}~\bibnamefont
  {Castin}},\ }\href {\doibase 10.1140/epjd/e2003-00015-y} {\bibfield
  {journal} {\bibinfo  {journal} {Eur. Phys. J. D}\ }\textbf {\bibinfo {volume}
  {22}},\ \bibinfo {pages} {235} (\bibinfo {year} {2003})}\BibitemShut
  {NoStop}%
\bibitem [{\citenamefont {Rosenbusch}\ \emph {et~al.}(2002)\citenamefont
  {Rosenbusch}, \citenamefont {Bretin},\ and\ \citenamefont
  {Dalibard}}]{Rosenbusch02}%
  \BibitemOpen
  \bibfield  {author} {\bibinfo {author} {\bibfnamefont {P.}~\bibnamefont
  {Rosenbusch}}, \bibinfo {author} {\bibfnamefont {V.}~\bibnamefont {Bretin}},
  \ and\ \bibinfo {author} {\bibfnamefont {J.}~\bibnamefont {Dalibard}},\
  }\href {\doibase 10.1103/PhysRevLett.89.200403} {\bibfield  {journal}
  {\bibinfo  {journal} {Phys. Rev. Lett.}\ }\textbf {\bibinfo {volume} {89}},\
  \bibinfo {pages} {200403} (\bibinfo {year} {2002})}\BibitemShut {NoStop}%
\bibitem [{\citenamefont {Lundh}\ and\ \citenamefont {Ao}(2000)}]{Lundh00}%
  \BibitemOpen
  \bibfield  {author} {\bibinfo {author} {\bibfnamefont {E.}~\bibnamefont
  {Lundh}}\ and\ \bibinfo {author} {\bibfnamefont {P.}~\bibnamefont {Ao}},\
  }\href {\doibase 10.1103/PhysRevA.61.063612} {\bibfield  {journal} {\bibinfo
  {journal} {Phys. Rev. A}\ }\textbf {\bibinfo {volume} {61}},\ \bibinfo
  {pages} {063612} (\bibinfo {year} {2000})}\BibitemShut {NoStop}%
\bibitem [{\citenamefont {Kim}\ and\ \citenamefont {Fetter}(2004)}]{Kim04}%
  \BibitemOpen
  \bibfield  {author} {\bibinfo {author} {\bibfnamefont {J.-k.}\ \bibnamefont
  {Kim}}\ and\ \bibinfo {author} {\bibfnamefont {A.}~\bibnamefont {Fetter}},\
  }\href {\doibase 10.1103/PhysRevA.70.043624} {\bibfield  {journal} {\bibinfo
  {journal} {Phys. Rev. A}\ }\textbf {\bibinfo {volume} {70}},\ \bibinfo
  {pages} {043624} (\bibinfo {year} {2004})}\BibitemShut {NoStop}%
\bibitem [{\citenamefont {Fedichev}\ and\ \citenamefont
  {Shlyapnikov}(1999)}]{Fedichev99}%
  \BibitemOpen
  \bibfield  {author} {\bibinfo {author} {\bibfnamefont {P.}~\bibnamefont
  {Fedichev}}\ and\ \bibinfo {author} {\bibfnamefont {G.}~\bibnamefont
  {Shlyapnikov}},\ }\href {\doibase 10.1103/PhysRevA.60.R1779} {\bibfield
  {journal} {\bibinfo  {journal} {Phys. Rev. A}\ }\textbf {\bibinfo {volume}
  {60}},\ \bibinfo {pages} {R1779} (\bibinfo {year} {1999})}\BibitemShut
  {NoStop}%
\bibitem [{\citenamefont {Yefsah}\ \emph {et~al.}(2013)\citenamefont {Yefsah},
  \citenamefont {Sommer}, \citenamefont {Ku}, \citenamefont {Cheuk},
  \citenamefont {Ji}, \citenamefont {Bakr},\ and\ \citenamefont
  {Zwierlein}}]{Yefsah13}%
  \BibitemOpen
  \bibfield  {author} {\bibinfo {author} {\bibfnamefont {T.}~\bibnamefont
  {Yefsah}}, \bibinfo {author} {\bibfnamefont {A.~T.}\ \bibnamefont {Sommer}},
  \bibinfo {author} {\bibfnamefont {M.~J.~H.}\ \bibnamefont {Ku}}, \bibinfo
  {author} {\bibfnamefont {L.~W.}\ \bibnamefont {Cheuk}}, \bibinfo {author}
  {\bibfnamefont {W.}~\bibnamefont {Ji}}, \bibinfo {author} {\bibfnamefont
  {W.~S.}\ \bibnamefont {Bakr}}, \ and\ \bibinfo {author} {\bibfnamefont
  {M.~W.}\ \bibnamefont {Zwierlein}},\ }\href@noop {} {\bibfield  {journal}
  {\bibinfo  {journal} {Nature}\ }\textbf {\bibinfo {volume} {499}},\ \bibinfo
  {pages} {426} (\bibinfo {year} {2013})}\BibitemShut {NoStop}%
\bibitem [{\citenamefont {Paoletti}\ \emph {et~al.}(2010)\citenamefont
  {Paoletti}, \citenamefont {Fisher},\ and\ \citenamefont
  {Lathrop}}]{Paoletti10}%
  \BibitemOpen
  \bibfield  {author} {\bibinfo {author} {\bibfnamefont {M.}~\bibnamefont
  {Paoletti}}, \bibinfo {author} {\bibfnamefont {M.~E.}\ \bibnamefont
  {Fisher}}, \ and\ \bibinfo {author} {\bibfnamefont {D.}~\bibnamefont
  {Lathrop}},\ }\href {\doibase http://dx.doi.org/10.1016/j.physd.2009.03.006}
  {\bibfield  {journal} {\bibinfo  {journal} {Physica D}\ }\textbf {\bibinfo
  {volume} {239}},\ \bibinfo {pages} {1367 } (\bibinfo {year}
  {2010})}\BibitemShut {NoStop}%
\bibitem [{\citenamefont {Allen}\ \emph {et~al.}(2014)\citenamefont {Allen},
  \citenamefont {Zuccher}, \citenamefont {Caliari}, \citenamefont {Proukakis},
  \citenamefont {Parker},\ and\ \citenamefont {Barenghi}}]{Allen14}%
  \BibitemOpen
  \bibfield  {author} {\bibinfo {author} {\bibfnamefont {A.}~\bibnamefont
  {Allen}}, \bibinfo {author} {\bibfnamefont {S.}~\bibnamefont {Zuccher}},
  \bibinfo {author} {\bibfnamefont {M.}~\bibnamefont {Caliari}}, \bibinfo
  {author} {\bibfnamefont {N.~P.}\ \bibnamefont {Proukakis}}, \bibinfo {author}
  {\bibfnamefont {N.~G.}\ \bibnamefont {Parker}}, \ and\ \bibinfo {author}
  {\bibfnamefont {C.~F.}\ \bibnamefont {Barenghi}},\ }\href {\doibase
  10.1103/PhysRevA.90.013601} {\bibfield  {journal} {\bibinfo  {journal} {Phys.
  Rev. A}\ }\textbf {\bibinfo {volume} {90}},\ \bibinfo {pages} {013601}
  (\bibinfo {year} {2014})}\BibitemShut {NoStop}%
\bibitem [{\citenamefont {Berry}\ and\ \citenamefont {Dennis}(2012)}]{Berry12}%
  \BibitemOpen
  \bibfield  {author} {\bibinfo {author} {\bibfnamefont {M.~V.}\ \bibnamefont
  {Berry}}\ and\ \bibinfo {author} {\bibfnamefont {M.~R.}\ \bibnamefont
  {Dennis}},\ }\href@noop {} {\bibfield  {journal} {\bibinfo  {journal} {Eur.
  J. Phys.}\ }\textbf {\bibinfo {volume} {33}},\ \bibinfo {pages} {723}
  (\bibinfo {year} {2012})}\BibitemShut {NoStop}%
\bibitem [{\citenamefont {Verbiest}\ and\ \citenamefont
  {Ach\'ucarro}(2011)}]{Verbiest11}%
  \BibitemOpen
  \bibfield  {author} {\bibinfo {author} {\bibfnamefont {G.~J.}\ \bibnamefont
  {Verbiest}}\ and\ \bibinfo {author} {\bibfnamefont {A.}~\bibnamefont
  {Ach\'ucarro}},\ }\href {\doibase 10.1103/PhysRevD.84.105036} {\bibfield
  {journal} {\bibinfo  {journal} {Phys. Rev. D}\ }\textbf {\bibinfo {volume}
  {84}},\ \bibinfo {pages} {105036} (\bibinfo {year} {2011})}\BibitemShut
  {NoStop}%
\bibitem [{\citenamefont {Leadbeater}\ \emph {et~al.}(2001)\citenamefont
  {Leadbeater}, \citenamefont {Winiecki}, \citenamefont {Samuels},
  \citenamefont {Barenghi},\ and\ \citenamefont {Adams}}]{Leadbeater01}%
  \BibitemOpen
  \bibfield  {author} {\bibinfo {author} {\bibfnamefont {M.}~\bibnamefont
  {Leadbeater}}, \bibinfo {author} {\bibfnamefont {T.}~\bibnamefont
  {Winiecki}}, \bibinfo {author} {\bibfnamefont {D.}~\bibnamefont {Samuels}},
  \bibinfo {author} {\bibfnamefont {C.}~\bibnamefont {Barenghi}}, \ and\
  \bibinfo {author} {\bibfnamefont {C.}~\bibnamefont {Adams}},\ }\href@noop {}
  {\bibfield  {journal} {\bibinfo  {journal} {Phys. Rev. Lett.}\ }\textbf
  {\bibinfo {volume} {86}},\ \bibinfo {pages} {1410} (\bibinfo {year}
  {2001})}\BibitemShut {NoStop}%
\bibitem [{\citenamefont {Vinen}(2005)}]{Vinen05}%
  \BibitemOpen
  \bibfield  {author} {\bibinfo {author} {\bibfnamefont {W.}~\bibnamefont
  {Vinen}},\ }\href@noop {} {\bibfield  {journal} {\bibinfo  {journal} {J.
  Phys.: Condens. Matter}\ }\textbf {\bibinfo {volume} {17}},\ \bibinfo {pages}
  {S3231} (\bibinfo {year} {2005})}\BibitemShut {NoStop}%
\bibitem [{\citenamefont {Fonda}\ \emph {et~al.}(2014)\citenamefont {Fonda},
  \citenamefont {Meichle}, \citenamefont {Ouellette}, \citenamefont {Hormoz},\
  and\ \citenamefont {Lathrop}}]{Fonda14}%
  \BibitemOpen
  \bibfield  {author} {\bibinfo {author} {\bibfnamefont {E.}~\bibnamefont
  {Fonda}}, \bibinfo {author} {\bibfnamefont {D.~P.}\ \bibnamefont {Meichle}},
  \bibinfo {author} {\bibfnamefont {N.~T.}\ \bibnamefont {Ouellette}}, \bibinfo
  {author} {\bibfnamefont {S.}~\bibnamefont {Hormoz}}, \ and\ \bibinfo {author}
  {\bibfnamefont {D.~P.}\ \bibnamefont {Lathrop}},\ }\href {\doibase
  10.1073/pnas.1312536110} {\bibfield  {journal} {\bibinfo  {journal} {Proc.
  Nat. Acad. Sci.}\ }\textbf {\bibinfo {volume} {111}},\ \bibinfo {pages}
  {4707} (\bibinfo {year} {2014})}\BibitemShut {NoStop}%
\bibitem [{\citenamefont {Sire}\ \emph {et~al.}(2011)\citenamefont {Sire},
  \citenamefont {Chavanis},\ and\ \citenamefont {Sopik}}]{Sire11}%
  \BibitemOpen
  \bibfield  {author} {\bibinfo {author} {\bibfnamefont {C.}~\bibnamefont
  {Sire}}, \bibinfo {author} {\bibfnamefont {P.-H.}\ \bibnamefont {Chavanis}},
  \ and\ \bibinfo {author} {\bibfnamefont {J.}~\bibnamefont {Sopik}},\ }\href
  {\doibase 10.1103/PhysRevE.84.056317} {\bibfield  {journal} {\bibinfo
  {journal} {Phys. Rev. E}\ }\textbf {\bibinfo {volume} {84}},\ \bibinfo
  {pages} {056317} (\bibinfo {year} {2011})}\BibitemShut {NoStop}%
\bibitem [{\citenamefont {Koplik}\ and\ \citenamefont
  {Levine}(1993)}]{Koplik93}%
  \BibitemOpen
  \bibfield  {author} {\bibinfo {author} {\bibfnamefont {J.}~\bibnamefont
  {Koplik}}\ and\ \bibinfo {author} {\bibfnamefont {H.}~\bibnamefont
  {Levine}},\ }\href {\doibase 10.1103/PhysRevLett.71.1375} {\bibfield
  {journal} {\bibinfo  {journal} {Phys. Rev. Lett.}\ }\textbf {\bibinfo
  {volume} {71}},\ \bibinfo {pages} {1375} (\bibinfo {year}
  {1993})}\BibitemShut {NoStop}%
\bibitem [{\citenamefont {Nazarenko}\ and\ \citenamefont
  {West}(2003)}]{Nazarenko03}%
  \BibitemOpen
  \bibfield  {author} {\bibinfo {author} {\bibfnamefont {S.}~\bibnamefont
  {Nazarenko}}\ and\ \bibinfo {author} {\bibfnamefont {R.}~\bibnamefont
  {West}},\ }\href@noop {} {\bibfield  {journal} {\bibinfo  {journal} {J. Low
  Temp. Phys.}\ }\textbf {\bibinfo {volume} {132}},\ \bibinfo {pages} {1}
  (\bibinfo {year} {2003})}\BibitemShut {NoStop}%
\bibitem [{\citenamefont {Gabbay}\ \emph {et~al.}(1998)\citenamefont {Gabbay},
  \citenamefont {Ott},\ and\ \citenamefont {Guzdar}}]{Gabbay98}%
  \BibitemOpen
  \bibfield  {author} {\bibinfo {author} {\bibfnamefont {M.}~\bibnamefont
  {Gabbay}}, \bibinfo {author} {\bibfnamefont {E.}~\bibnamefont {Ott}}, \ and\
  \bibinfo {author} {\bibfnamefont {P.}~\bibnamefont {Guzdar}},\ }\href
  {\doibase 10.1103/PhysRevE.58.2576} {\bibfield  {journal} {\bibinfo
  {journal} {Phys. Rev. E}\ }\textbf {\bibinfo {volume} {58}},\ \bibinfo
  {pages} {2576} (\bibinfo {year} {1998})}\BibitemShut {NoStop}%
\bibitem [{\citenamefont {Wells}\ \emph {et~al.}(2015)\citenamefont {Wells},
  \citenamefont {Lode}, \citenamefont {Bagnato},\ and\ \citenamefont
  {Tsatsos}}]{Wells15}%
  \BibitemOpen
  \bibfield  {author} {\bibinfo {author} {\bibfnamefont {T.}~\bibnamefont
  {Wells}}, \bibinfo {author} {\bibfnamefont {A.}~\bibnamefont {Lode}},
  \bibinfo {author} {\bibfnamefont {V.}~\bibnamefont {Bagnato}}, \ and\
  \bibinfo {author} {\bibfnamefont {M.}~\bibnamefont {Tsatsos}},\ }\href
  {\doibase 10.1007/s10909-015-1285-y} {\bibfield  {journal} {\bibinfo
  {journal} {Journal of Low Temperature Physics}\ }\textbf {\bibinfo {volume}
  {180}},\ \bibinfo {pages} {133} (\bibinfo {year} {2015})}\BibitemShut
  {NoStop}%
\bibitem [{\citenamefont {Henn}\ \emph {et~al.}(2009)\citenamefont {Henn},
  \citenamefont {Seman}, \citenamefont {Roati}, \citenamefont {Magalh{\~a}es},\
  and\ \citenamefont {Bagnato}}]{Henn09}%
  \BibitemOpen
  \bibfield  {author} {\bibinfo {author} {\bibfnamefont {E.~A.~L.}\
  \bibnamefont {Henn}}, \bibinfo {author} {\bibfnamefont {J.~A.}\ \bibnamefont
  {Seman}}, \bibinfo {author} {\bibfnamefont {G.}~\bibnamefont {Roati}},
  \bibinfo {author} {\bibfnamefont {K.~M.~F.}\ \bibnamefont {Magalh{\~a}es}}, \
  and\ \bibinfo {author} {\bibfnamefont {V.~S.}\ \bibnamefont {Bagnato}},\
  }\href {\doibase 10.1103/PhysRevLett.103.045301} {\bibfield  {journal}
  {\bibinfo  {journal} {Phys. Rev. Lett.}\ }\textbf {\bibinfo {volume} {103}},\
  \bibinfo {pages} {045301} (\bibinfo {year} {2009})}\BibitemShut {NoStop}%
\bibitem [{\citenamefont {Seman}\ \emph {et~al.}(2010)\citenamefont {Seman},
  \citenamefont {Henn}, \citenamefont {Haque}, \citenamefont {Shiozaki},
  \citenamefont {Ramos}, \citenamefont {Caracanhas}, \citenamefont {Castilho},
  \citenamefont {Castelo~Branco}, \citenamefont {Tavares}, \citenamefont
  {Poveda-Cuevas}, \citenamefont {Roati}, \citenamefont {Magalh{\~a}es},\ and\
  \citenamefont {Bagnato}}]{Seman10}%
  \BibitemOpen
  \bibfield  {author} {\bibinfo {author} {\bibfnamefont {J.}~\bibnamefont
  {Seman}}, \bibinfo {author} {\bibfnamefont {E.}~\bibnamefont {Henn}},
  \bibinfo {author} {\bibfnamefont {M.}~\bibnamefont {Haque}}, \bibinfo
  {author} {\bibfnamefont {R.}~\bibnamefont {Shiozaki}}, \bibinfo {author}
  {\bibfnamefont {E.}~\bibnamefont {Ramos}}, \bibinfo {author} {\bibfnamefont
  {M.}~\bibnamefont {Caracanhas}}, \bibinfo {author} {\bibfnamefont
  {P.}~\bibnamefont {Castilho}}, \bibinfo {author} {\bibfnamefont
  {C.}~\bibnamefont {Castelo~Branco}}, \bibinfo {author} {\bibfnamefont
  {P.}~\bibnamefont {Tavares}}, \bibinfo {author} {\bibfnamefont
  {F.}~\bibnamefont {Poveda-Cuevas}}, \bibinfo {author} {\bibfnamefont
  {G.}~\bibnamefont {Roati}}, \bibinfo {author} {\bibfnamefont
  {K.}~\bibnamefont {Magalh{\~a}es}}, \ and\ \bibinfo {author} {\bibfnamefont
  {V.}~\bibnamefont {Bagnato}},\ }\href@noop {} {\bibfield  {journal} {\bibinfo
   {journal} {Phys. Rev. A}\ }\textbf {\bibinfo {volume} {82}},\ \bibinfo
  {pages} {033616} (\bibinfo {year} {2010})}\BibitemShut {NoStop}%
\bibitem [{\citenamefont {Yukalov}\ \emph {et~al.}(2015)\citenamefont
  {Yukalov}, \citenamefont {Novikov},\ and\ \citenamefont
  {Bagnato}}]{Yukalov15}%
  \BibitemOpen
  \bibfield  {author} {\bibinfo {author} {\bibfnamefont {V.}~\bibnamefont
  {Yukalov}}, \bibinfo {author} {\bibfnamefont {A.}~\bibnamefont {Novikov}}, \
  and\ \bibinfo {author} {\bibfnamefont {V.}~\bibnamefont {Bagnato}},\ }\href
  {\doibase 10.1007/s10909-015-1288-8} {\bibfield  {journal} {\bibinfo
  {journal} {Journal of Low Temperature Physics}\ }\textbf {\bibinfo {volume}
  {180}},\ \bibinfo {pages} {53} (\bibinfo {year} {2015})}\BibitemShut
  {NoStop}%
\end{thebibliography}

\begin{thebibliography}{1}
\expandafter\ifx\csname natexlab\endcsname\relax\def\natexlab#1{#1}\fi
\expandafter\ifx\csname bibnamefont\endcsname\relax
  \def\bibnamefont#1{#1}\fi
\expandafter\ifx\csname bibfnamefont\endcsname\relax
  \def\bibfnamefont#1{#1}\fi
\expandafter\ifx\csname citenamefont\endcsname\relax
  \def\citenamefont#1{#1}\fi
\expandafter\ifx\csname url\endcsname\relax
  \def\url#1{\texttt{#1}}\fi
\expandafter\ifx\csname urlprefix\endcsname\relax\def\urlprefix{URL }\fi
\providecommand{\bibinfo}[2]{#2}
\providecommand{\eprint}[2][]{\url{#2}}

\bibitem[{\citenamefont{Zobay and Garraway}(2001)}]{S_Zobay01}
\bibinfo{author}{\bibfnamefont{O.}~\bibnamefont{Zobay}} \bibnamefont{and}
  \bibinfo{author}{\bibfnamefont{B.~M.} \bibnamefont{Garraway}},
  \bibinfo{journal}{Phys. Rev. Lett.} \textbf{\bibinfo{volume}{86}},
  \bibinfo{pages}{1195} (\bibinfo{year}{2001}).

\end{thebibliography}

%merlin.mbs apsrev4-1.bst 2010-07-25 4.21a (PWD, AO, DPC) hacked"

%%%%%%%%%% Merge with supplemental materials %%%%%%%%%%
\pagebreak
\widetext
\begin{center}
\textbf{\large SUPPLEMENTAL MATERIAL}
\end{center}
%%%%%%%%%% Merge with supplemental materials %%%%%%%%%%
%%%%%%%%%% Prefix a "S" to all equations, figures, tables and reset the counter %%%%%%%%%%
\setcounter{equation}{0}
\setcounter{figure}{0}
\setcounter{table}{0}
\setcounter{page}{1}
\makeatletter
\renewcommand{\theequation}{S\arabic{equation}}
\renewcommand{\thefigure}{S\arabic{figure}}
\renewcommand{\bibnumfmt}[1]{[S#1]}
\renewcommand{\citenumfont}[1]{S#1}
%%%%%%%%%% Prefix a "S" to all equations, figures, tables and reset the counter %%%%%%%%%%

\section{Expansion of the outcoupled atoms}
During the expansion of the outcoupled atoms, optical levitation is performed with a blue-detuned $532$ nm laser beam to compensate for gravity and a radio frequency dressing~\cite{S_Zobay01} is used to keep the out-coupled fraction confined and clearly detectable after the $13$ ms expansion. In particular, the RF field is such to produce a mexican-hat potential which limits the radial expansion to about $100$ $\mu$m, whereas the slower axial expansion is barely perturbed.

\section{Vortex oscillations}

\begin{figure}[!ht]       %--------------FIGURE 1--------------
\centering
\includegraphics[width=1.0\columnwidth]{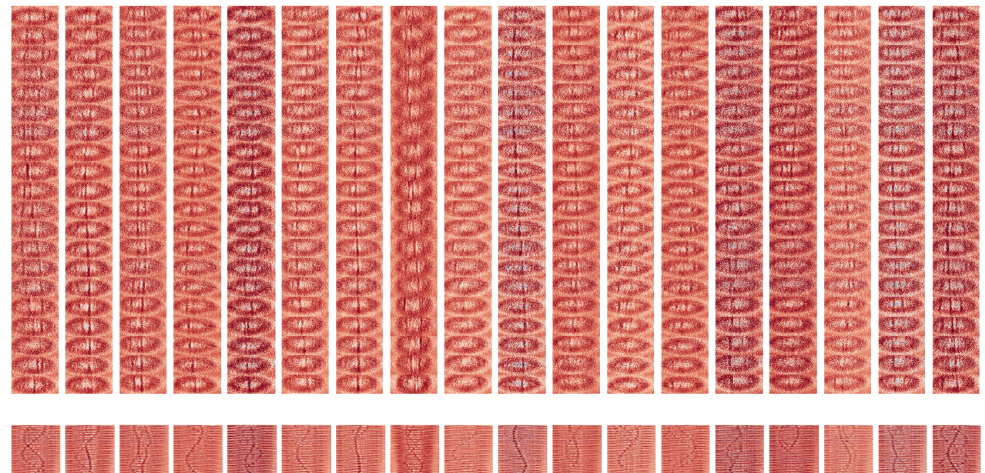}
\caption{Examples of experimental images taken with the stroboscopic outcoupling technique, reported in real scale (top) and squeezed in order to improve defects visibility (bottom).}
\label{osc}
\end{figure}

\begin{figure}[!ht]       %--------------FIGURE 1--------------
\centering
\includegraphics[width=0.5\columnwidth]{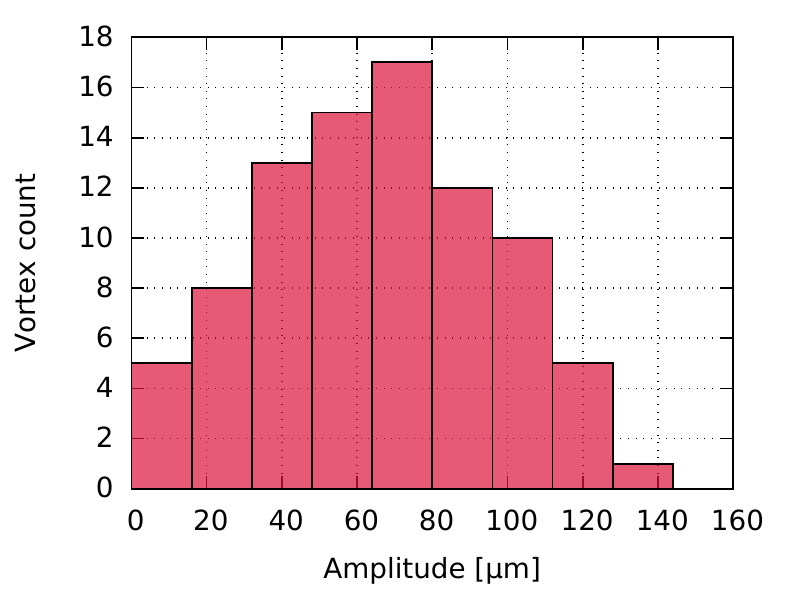}
\caption{Occurence of amplitude in the vortex oscillations after expansion, the axial TF radius after expansion is $\sim 250$ $\mu$m. The in-situ value can be obtained considering a scale factor of $\sim 0.6$, given by the ratio between the in-situ and expanded TF radius at $t=0$ ms; this because the assumption of a constant $r_0$ during the expansion. This gives a mean $r_o$ of $0.27$ with a standard deviation of $0.13$. There is no statistical difference between the single-vortex distribution and the double-vortex one.}
\label{amplHist}
\end{figure}

\newpage
\section{Phase shift and relative velocity}
A precise statistical analysis is not possible here because information on the phase shift can be extracted only in the data subset where the crossing point occurs at about half of the inspected time evolution ($\sim10\%$ of the cases). Clear phase shifts are present in about half of this subset.

\begin{figure}[!h]       %--------------FIGURE 2--------------
\centering
\includegraphics[width=0.5\columnwidth]{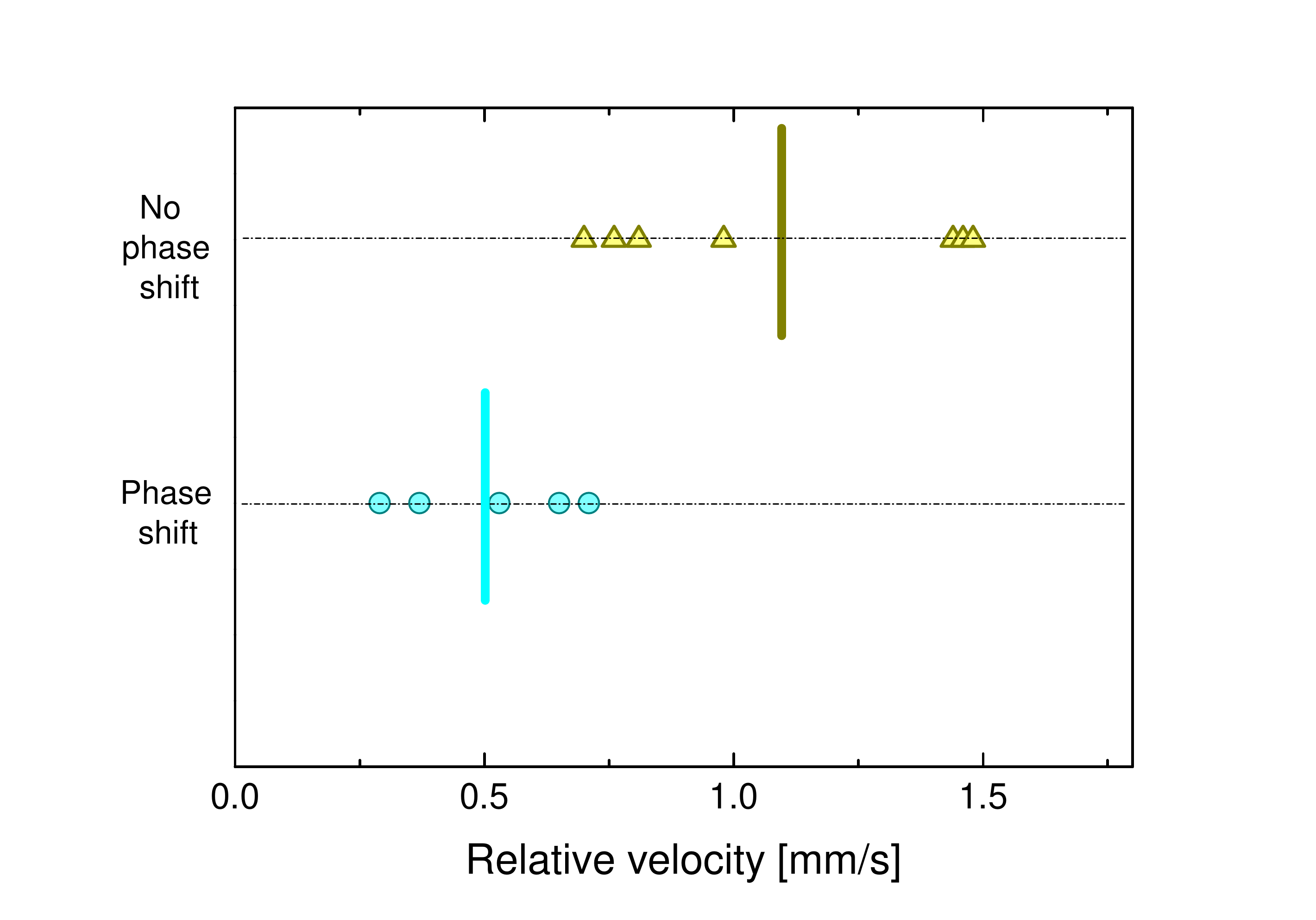}
\caption{Relative velocity between vortices whose crossing trajectories clearly show or not a phase shift. Velocities are calculated differentiating the function fitting the vortex trajectories at the crossing point and rescaled to take into account expansion; vertical lines represent the mean velocities in the two cases.}
\label{RelVel}
\end{figure}

\end{document}